\documentclass[]{cupconf}
\usepackage{graphicx}

\newcommand{\oversim}[2]{\protect{\mbox{\lower0.5ex\vbox{%
   \baselineskip=0pt\lineskip=0.2ex
   \ialign{$\mathsurround=0pt #1\hfil##\hfil$\crcr#2\crcr\sim\crcr}}}}} 
\newcommand{\simgreat}{\mbox{$\,\mathrel{\mathpalette\oversim>}\,$}} 
\newcommand{\simless} {\mbox{$\,\mathrel{\mathpalette\oversim<}\,$}} 

\title[The metal-rich IMF]
      {The stellar initial mass function of metal-rich populations
\footnote{\small To appear in 
{\it The Metal Rich Universe}, La Palma, Canary Islands, June 12th--16th, 2006, 
G. Israelian \& G. Meynet (eds.), Cambridge University Press}
}

\author[P. Kroupa]{\\ Pavel Kroupa} 
\affiliation{Argelander Institute for Astronomy, University of Bonn, Auf dem H\"ugel 71,
D-53121 Bonn, Germany}

\begin{document}
\maketitle

\begin{abstract}
Does the IMF vary? Is it significantly different in metal-rich
environments than in metal-poor ones? Theoretical work predicts this
to be the case.  But in order to provide robust empirical evidence
for
 this, the researcher {\it must} understand {\it all} possible
biases
 affecting the derivation of the stellar mass function. Apart
from the very difficult observational challenges, this turns out to be
highly non-trivial relying on an exact understanding of how {\it stars
evolve}, and how stellar populations in galaxies are assembled
dynamically and how individual star clusters and associations
evolve. {\it $N$-body modelling} is therefore an unavoidable tool in
this game: the case can be made that without complete dynamical
modelling of star clusters and associations any statements about the
variation of the IMF with physical conditions are most probably
wrong. The calculations
 that do exist demonstrate time and again
that the IMF is invariant:
 There exists no statistically meaningful
evidence for a variation of
 the IMF from metal-poor to metal-rich
populations. This means that
 currently existing star-formation
theory fails to describe the stellar outcome.  Indirect evidence,
based on chemical evolution calculations, however indicate that
extreme star-bursts that assembled bulges and elliptical galaxies
may
 have had a top-heavy IMF.
\end{abstract}

\firstsection
\section{Introduction}
\label{sec_kroupa:intro}

\noindent
 The stellar initial mass function (IMF), $\xi(m)\,dm$,
where $m$ is
 the stellar mass, is the parent distribution function
of the masses of stars formed in {\it one}
 event. Here, the number
of stars in the mass interval $m,m+dm$ is $dN
 = \xi(m)\,dm$.
Salpeter (1955) inferred the IMF from
 solar-neighbourhood
star-counts applying corrections for stellar
 evolution and
Galactic-disk structure finding $\xi(m)\approx
 k\,m^{-\alpha},
\alpha \approx 2.35$, for $0.4 \simless m/M_\odot
 \simless
10$. Miller \& Scalo (1979) and Scalo (1986) derived the IMF for
$0.1-60\,M_\odot$ using better data and a more sophisticated analysis
establishing that
 the IMF flattens or turns-over at small masses.
Modern studies of
 solar-neighbourhood star-count data which also
apply detailed
 corrections for unknown multiple systems in the
star-counts, confirm
 that $\alpha_2=2.2\pm0.3$ for $0.5 \simless
m/M_\odot\simless 1$
 (Kroupa, Tout \& Gilmore 1993; Kroupa 1995b;
Reid, Gizis \& Hawley
 2002). Using spectroscopic star-by-star
observations, Massey (2003)
 reports the same {\it slope or index}
$\alpha_2=2.3\pm0.1$ for $m\simgreat 10\,M_\odot$ in many OB
associations and star clusters in the Milky Way (MW), the Large- and
Small-Magellanic clouds (LMC, SMC, respectively). It is therefore
suggested to refer to $\alpha_2=2.3$ as the {\it Salpeter/Massey
slope}
 or {\it index}. It is valid for $m\simgreat 0.5\,M_\odot$.

The IMF is, strictly speaking, a hypothetical construct because any
{\it observed} system of $N$ stars merely constitutes a particular
{\it representation} of a distribution function. The probable
existence of a unique $\xi(m)$ can be inferred from observations of
many ensembles of such $N$ systems (e.g. Massey 2003). If, after
corrections for

(a) stellar evolution, 

(b) unknown multiple stellar systems and 

(c) stellar-dynamical biases, 

\noindent the individual distributions of stellar masses is similar
{\it within the statistical uncertainties}, then we (the community)
deduce that the hypothesis that the stellar mass distributions are
not
 the same can be excluded. That is, we make the case for a {\it
universal, standard} or {\it canonical} {\it stellar IMF} within the
physical conditions probed by the relevant physical parameters
(metallicity, density, mass) of the populations at hand.

This canonical IMF is a two-part-power law, the only structure with
confidence found so far being the change of index from the
Salpeter/Massey value to a smaller one near $0.5\,M_\odot$:
\begin{equation}
\begin{array}{l@{\quad\quad,\quad}r@{\;}l}
\alpha_1 = 1.3\pm0.5,    &0.08 \simless &m/M_\odot \simless 0.5,\\
\alpha_2 = 2.3\pm0.5,    &0.5 \simless  &m/M_\odot \simless 150.
\end{array}
\label{eq_kroupa:canonIMF}
\end{equation}
It has been fully corrected for unknown multiple stellar systems in
the low-mass ($m < 1\,M_\odot$) regime, while multiplicity
corrections
 in the high-mass regime await to be done. The evidence
for a universal
 upper mass cutoff near $150\,M_\odot$ (Weidner \&
Kroupa 2004; Figer
 2005; Oey \& Clarke 2005; Koen 2006) seems to be
rather well
 established in populations with metallicities ranging
from the LMC
 ($Z\approx 0.008$) to the super-solar Galactic centre
($Z\simgreat
 0.02$) such that the stellar mass function (MF) simply
stops at that mass. This mass needs to be understood theoretically
(see discussion in Kroupa \& Weidner 2005).

Chabrier (2003) offers a log-normal form\footnote{\small Note that the
log-normal form is physically better motivated in the sense that a
physical process will not abruptly change a slope as in the
canonical
 IMF, but it also needs to be extended by a power-law
above
 $1\,M_\odot$ to meet the needs of the observational data
thereby
 losing its advantage. A reason why the author prefers to use
the
 canonical form is mathematical simplicity and the ease with
which
 parts of it can be changed without affecting other parts.}
which fits
 the canonical form quite well (e.g. Romano et al. 2005).

Below the hydrogen-burning limit, there is substantial evidence that
the IMF flattens further to $\alpha\approx 0.3\pm0.7$ (Kroupa 2001a;
Kroupa 2002; Chabrier 2003).  Therefore, the canonical IMF most
likely
 has a peak at $0.08\,M_\odot$. Brown dwarfs, however,
comprise only a
 few~\% of the mass of a population and are therefore
dynamically
 irrelevant.  Note that the logarithmic form of the
canonical IMF,
 $\xi_{\rm L}(m) = {\rm ln}(10)\;m\;\xi(m)$, which
gives the number of
 stars in log$_{10}m$-intervals, also has a peak
near $0.08\,M_\odot$. However, the {\it system} IMF (of stellar
companions per binary combined to system masses) has a maximum in the
mass range $0.4-0.6\,M_\odot$ (Kroupa et al. 2003).

The above form has been derived from detailed considerations of
local
 star-counts thereby representing an {\it average} IMF: for
low-mass
 stars it is a mixture of stellar populations spanning a
large range of
 ages ($10-0$~Gyr) and metallicities ([Fe/H]$\simless
0$), while for
 the massive stars it constitutes a mixture of
different metallicities
 ([Fe/H]$\simgreat -1.5$) and star-forming
conditions (OB associations
 to very dense star-burst clusters: R136
in the LMC). Therefore it can
 be taken to be a canonical form, and
the aim is to test whether even
 more extreme star-forming conditions
such as found in super-metal rich
 environments or super-dense
regions may deviate from it. Any {\it
 systematic} deviations of the
IMF with physical conditions of the
 environment would constrain our
understanding of star formation,
 would give us a prescription of how
to set-up stellar-dynamical
 systems, and last not least would allow
more precise galaxy-formation and evolution calculations.

\section{The expectation: the IMF must depend on star-formation environment and in 
particular on the metallicity}
\label{sec_kroupa:theory}

\noindent There are two basic arguments suggesting that the IMF
ought to be dependent on the physical conditions of star
formation:

\subsection{The Jeans-mass argument}

\noindent A) A region of a molecular cloud undergoing gravitational
collapse will have over-dense sub-regions within it which are also
Jeans-unstable collapsing independently to form smaller structures
which themselves may again sub-fragment (e.g. Zinnecker 1984).
Ultimately stars result. The essence of this concept is that a region 
spanning a Jeans length which has at least a Jeans mass
undergoes gravitational collapse. The Jeans mass depends on
the temperature and density of the cloud, $M_{\rm Jeans}\propto
T^{3/2}\;\rho^{-1/2}$ (e.g. Larson 1998; Bonnell, Larson \&
Zinnecker 2006). Now, in metal-rich
 environments there is more dust
and therefore the collapsing gas can
 cool more effectively reducing
$T$ and increasing $\rho$. Thus,
\begin{equation}
[{\rm Fe/H}] \; \uparrow \quad \Longrightarrow \quad {\rm
fragment\;masses}\; \downarrow\;.
\end{equation}

B) The fact that the IMF is not a featureless power-law but 
has structure in the mass range $0.08\simless m/M_\odot \simless 0.5$
suggests there to be a characteristic mass of a few~$0.1\,M_\odot$.

Bonnell et al. (2006) suggest that this characteristic mass of
fragmentation may be a result of the coupling of gas to dust such
that
 there is a change from a cooling equation of state, where
$T\propto
 \rho^{-0.25}$ while the density increases, to one with
slight heating at
 high densities, $T\propto \rho^{0.1}$. Again, this
implies a
 dependency of the characteristic mass on the metallicity
through the cooling rate:
\begin{equation}
[{\rm Fe/H}] \; \uparrow \quad \Longrightarrow \quad {\rm
fragment\;masses}\; \downarrow\;.
\end{equation}

There seems to be observational evidence supporting the notion that
stellar masses are derived from Jeans-unstable mass fragments:
pre-stellar cloud cores are found to be distributed like the canonical
IMF in low-mass star formating regions (Motte, Andre \& Neri 1998;
Testi \& Sargent 1998; Motte et al. 2001; but see Nutter \&
Ward-Thompson 2006).

While the concept of the Jeans mass is very natural and allows one
to
 nicely visualise the physical process of fragmentation, it has
the
 problem that the densest regions of a pre-star-cluster cloud
core
 ought to have the smaller fragment masses, but instead the
most
 massive stars are seen to form in the densest regions.

\subsection{The self-limitation argument}

\noindent A rather convincing physical model of the IMF which avoids
the problem
 with the Jeans mass argument has been suggested by Adams
\& Fatuzzo
 (1996) and Adams \& Laughlin (1996). The argument here is
that the
 Jeans mass has virtually nothing to do with the final mass
of a star
 because structure in a molecular cloud exists on all
scales. Therefore, no characteristic density can be identified, and
``no single Jeans mass exists''. When a cloud region becomes
unstable
 a hydrostatic core forms after the initial
free-collapse. This core
 then accretes at a rate dictated by the
physical conditions in the
 cloud. The metallicity influences the
accretion rate through the sound
 velocity (higher sound velocity,
larger accretion rate) by steering
 the cooling rate (more metals
$\Rightarrow$ more dust $\Rightarrow$
 lower $T$ $\Rightarrow$
smaller sound velocity). The many physical variables describing the
formation of a
 single star have distributions, and folding these
together yields
 finally an IMF in broad agreement with a log-normal
shape (as also
 shown by Zinnecker 1984). This fails though at large
masses, where the IMF is a power-law and where additional physical
processes probably play a role (coagulation, competitive accretion).
The important point however is that this theory {\it also} 
 expects
a variation of the resulting characteristic mass with metallicity as
above:
\begin{equation}
[{\rm Fe/H}] \; \uparrow \quad \Longrightarrow \quad {\rm
characteristic\;mass}\; \downarrow\;.
\end{equation}

\subsection{Robust implication?}

\noindent Thus, both theoretical lines of argument seem to
suggest the same
 qualitative behaviour, namely that the IMF ought to
shift to smaller
 characteristic masses with increasing
metallicity. This would
 therefore seem to suggest a {\it very
robust} if not {\it fundamental}
 expectation of star-formation
theory.

Is it born out by empirical
 evidence?  The best way to test this
theoretical expectation is to
 measure the IMF in metal-poor
environments and to compare to the
 shape seen in metal-rich
environments. 

A measurement of the IMF in a
 population with super-solar abundance
would be especially
 important. Unfortunately this is extremely
difficult, because
 populations with super-solar abundances are
exceedingly rare. Only one
 star-cluster with [Fe/H]$\approx+0.04$ is
known, NGC~6791, but its
 ancient age of $\approx 10$~Gyr implies it
to be dynamically very
 old. Evaporation of low-mass stars will thus
have significantly
 affected the shape of the present-day mass
function, which has not
 been measured yet (King et al. 2005). Other
metal-rich environments
 constitute the central MW region
(\S~\ref{sec_kroupa:GalC}) and galactic spheroids
(\S~\ref{sec_kroupa:bulge}), the latter
 allowing only indirect
evidence on the IMF through their chemical
 properties.

In general, measurements of the IMF are hard, because stellar masses
can only be inferred indirectly through their luminosity, $l$, which
also depends on age, metallicity and the star's spin angular momentum
vector, collectively producing a distribution of $l$ for a given $m$.
The empirical knowledge that can be gained about the IMF is discussed
next.

\section{The shape of the IMF}

\subsection{The local stellar sample}

\noindent
The best knowledge about the IMF we glean from the local
volume-limited stellar sample.  The basic technique is to construct
the stellar luminosity function (LF), $\Psi(M_P)$, where $M_P$ is the
stellar absolute magnitude in some photometric pass band (we are stuck
with magnitudes rather than working with the physically more intuitive
luminosities due to our historical inheritance). The number of stars
within the complete volume in the magnitude interval $M_P, M_P+dM_P$
is then $dN=\Psi(M_P)\,dM_P$. These are the same stars as enter
$dN=\xi(m)\,dm$ from above, and thus results our master equation
\begin{equation}
\Psi(M_P)=-{dm\over M_P}\;\xi(m).
\label{eq_kroupa:lf_mf}
\end{equation}
The observable is $\Psi$ which we get from the sky. Our target is
$\xi$, and the hurdle is the {\it derivative of
 the stellar
mass--luminosity relation}, $dm/dM_P$. This is quite
 problematical,
because we can only get at $dm/dM_P$ by either constructing
observational mass--luminosity relations using extremely
well-observed
 binary stars with known Kepler solutions which
nevertheless have
 uncertainties that are magnified when considering
the derivative, or
 we can resort to theoretical stellar models which
give us well defined
 derivatives but depend on theoretically
difficult processes within
 stellar interiors (opacities, convection,
rotation, magnetic fields,
 equation of state, nuclear energy
generation processes).

There are two basic local LF's: (I) We can count all stars within a
trigonometrically-defined distance limit such that the stellar
sample
 is complete, i.e. we can see {\it all} stars of magnitude
$M_P,M_P+dMP$ within a distance $r_t$. The volume-limited sample for
solar-type stars having excellent parallax measurements extends to
$r_t\approx20-30$~pc, while for the
 faintest M-dwarfs
$r_t\approx5-8$~pc (Reid, Gizis \& Hawley
 2002). Tests of
completeness are made by comparing the stars with
 $M_P$ within $r_t$
to the number of these stars in a volume element
 further out (Henry
et al. 2002), finding that faint stars remain to be
 discovered even
within a distance of 5~pc. For this {\it nearby LF},
 $\Psi_{\rm
neraby}$, the stars are well scrutinised on an
 individual-object
basis, and geometric distances are known to within
 about 10~per
cent. At the faint end $\Psi_{\rm nearby}$ is badly constrained 
resting on only a few stars.

(II) Prompted by the ``discovery'' of large amounts of dark matter in
the MW disk (Bahcall 1984)\footnote{The evidence for dark matter
within the solar vicinity disappeared on closer scrutiny (Kuijken \&
Gilmore 1991).}, novel deep surveys were pioneered by Reid \& Gilmore
(1982).  This second-type of sampling can be obtained by performing
deep, pencil-beam photographic or CCD imaging surveys through the
Galactic
 disk. From the $10^5$ images the typically 100 or so main
sequence
 stars need to be gleaned using automatic image-, colour and
brightness
 recognition systems. The distances of the stars are
determined using
 the method of {\it photometric parallax}, which
relies on estimating
 the absolute luminosity of a star from its
colour, and then
 calculating its distance from the distance
modulus. The resulting
 flux-limited sample of stars has
photometric-distance limits within
 which the counts are
complete. The distance limits decrease for fainter stars.

Clearly, while only {\it one nearby LF} exists, {\it many
photometric
 LFs} can be constructed for different fields of view.
Each
 observation yields a few dozen to hundreds of stars, and so
the
 overall sample size becomes very significant. The various
surveys have
 shown $\Psi_{\rm phot}$ to be invariant with
direction. This should, of course, be the case, since the
Galactic-field stars with an average age of about 5~Gyr have a
velocity dispersion of about 25-50~pc/Myr such that within 200~Myr a
volume with a dimension of the survey volumes (a few hundred~pc) is
completely mixed.

\begin{figure}
\begin{center}
\rotatebox{0}{\resizebox{0.8
\textwidth}{!}{\includegraphics{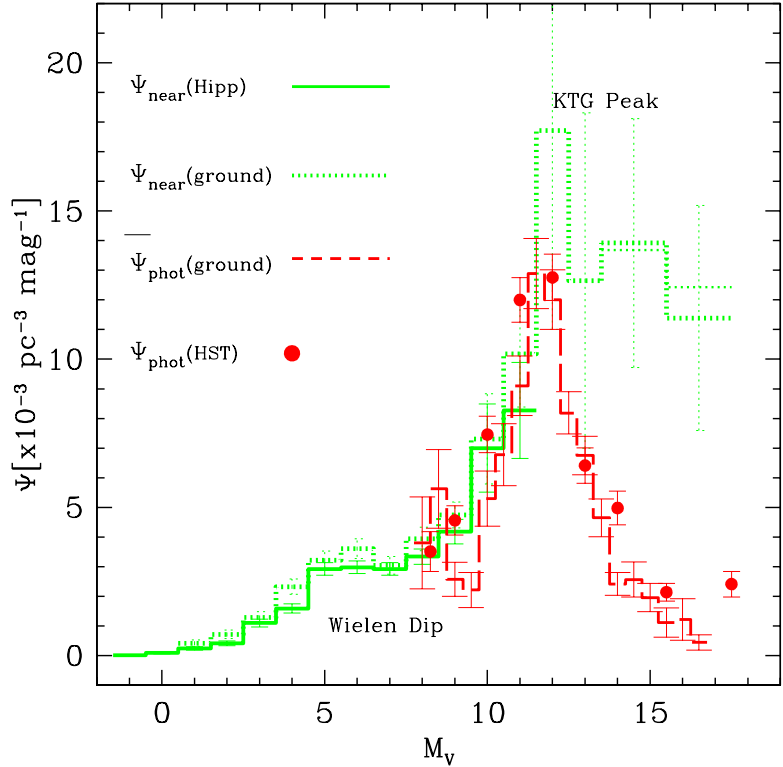}}}
\vskip -25mm
\caption{\small{Stellar luminosity functions (LFs) for
solar-neighbourhood stars. The photometric LF corrected for Malmquist
bias and at the mid-plane of the Galactic disk ($\Psi_{\rm phot}$) is
compared with the nearby LF ($\Psi_{\rm near}$). The average,
ground-based $\overline{\Psi}_{\rm phot}$ (dashed histogram, data
pre-dating 1995, Kroupa 1995a) is confirmed by Hubble-Space-Telescope
(HST) star-count data which pass through the entire Galactic disk and
are thus less prone to Malmquist bias (solid dots, Zheng et al. 2001).
The ground-based volume-limited trigonometric-parallax sample (dotted
histogram) systematically overestimates $\Psi_{\rm near}$ due to the
Lutz-Kelker bias, thus lying above the improved estimate provided by
the Hipparcos-satellite data (solid histogram, Jahrei{\ss} \& Wielen
1997; Kroupa 2001b).  The depression/plateau near $M_V=7$ is the {\it
Wielen dip}.  The maximum near $M_V\approx 12, M_I\approx 9$ is the
{\it KTG peak}. The thin dotted histogram at the faint end indicates
the level of refinement provided by recent stellar additions (Kroupa
2001b) demonstrating that even the immediate neighbourhood within
5.2~pc of the Sun probably remains incomplete at the faintest stellar
luminosities. }}
\label{fig_kroupa:MWlf}
\end{center}
\end{figure}

It therefore came as a surprise that $\Psi_{\rm
near}$ and $\Psi_{\rm
 phot}$ are significantly different at faint
luminosities
 (Fig.~\ref{fig_kroupa:MWlf}).

\subsection{The mass--luminosity relation}

\noindent
 When not understanding something the best strategy to
continue is
 sometimes to simply ``forget'' the problem and continue
the path of
 least resistance. Thus, while the problem $\Psi_{\rm
near} \ne
 \Psi_{\rm phot}, M_V>12$ could not be explained
immediately, it turned
 out to be constructive to first ascertain
which LF shape must be the
 correct one using entirely different
arguments. 

In eq.~\ref{eq_kroupa:lf_mf} the slope of the stellar
mass--luminosity
 relation of stars enters posing a clue. Fig.~2 in
Kroupa (2002) shows
 the mass--luminosity data of binary stars with
Kepler orbits, and
 demonstrates that a non-linearity exists near
$0.33\,M_\odot$ such
 that a pronounced peak in $-dm/dM_V$ appears at
$M_V\approx 11.5$ with
 an amplitude and width essentially identical
to the maximum seen in the photometric LF at
 this luminosity
(Fig.~\ref{fig_kroupa:MWlf}). This agreement of 
\begin{enumerate}
\item the location of the maximum {\it and} 
\item the amplitude {\it and}
\item the width of the extremum 
\end{enumerate}
convincingly suggest stellar astrophysics to be the origin of the
peak
 in the LF, rather than the MF. The Wielen dip
(Fig.~\ref{fig_kroupa:MWlf}) similarly results from subtle structure
in the mass--luminosity relation.  {\it Thus, simply by counting stars
on
 the sky we are able to direct our gaze within their interiors}:
It is
 the internal constitution of stars which changes with
changing
 main-sequence mass and this is what drives the structure in
the mass--luminosity relation.

\begin{figure}
\begin{center}
\rotatebox{0}{\resizebox{0.8
\textwidth}{!}{\includegraphics{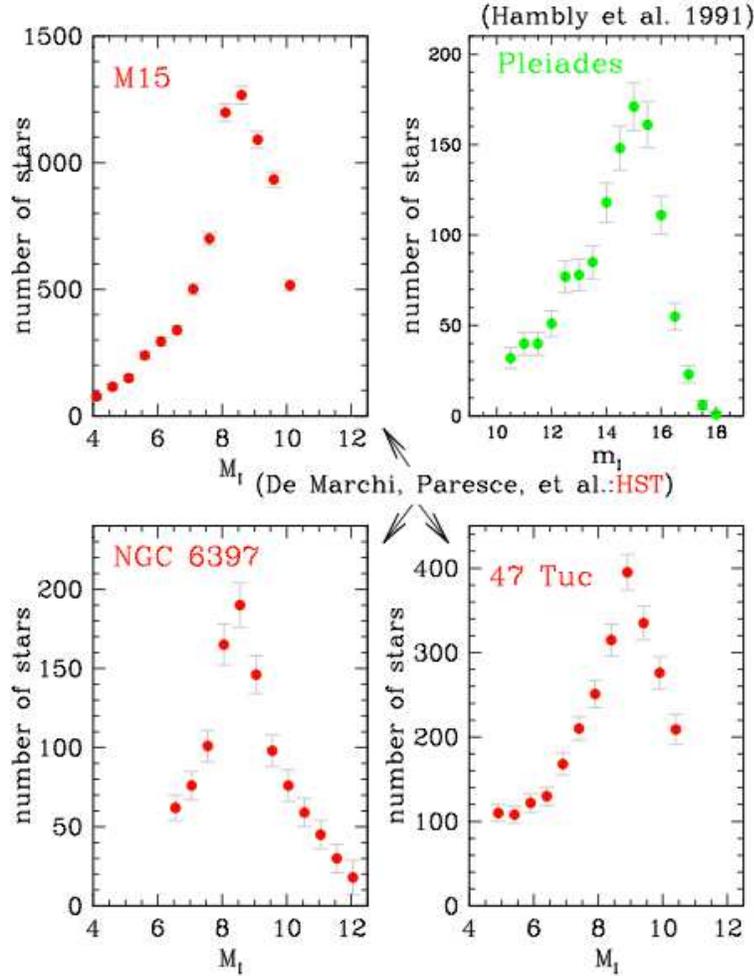}}}
\vskip 5mm
\caption{\small{$I$-band LFs of stellar {\it systems} in four star
clusters: globular cluster (GC) {\it M15} (de Marchi \& Paresce 1995a, distance
modulus $\Delta m=m-M=15.25$~mag); GC {\it NGC~6397}
(Paresce et al. 1995, $\Delta m=12.2$); young Galactic cluster {\it
Pleiades} (Hambly et al. 1991, $\Delta m=5.48$); GC {\it 47~Tuc}
(de Marchi et al. 1995b, $\Delta m=13.35$).
 }}
\label{fig_kroupa:CLlf}
\end{center}
\end{figure}

Having thus established that the peak in the LF {\it must, in fact} be
there where it is also found as a result of fundamental astrophysical
processes, we can test this result using star-clusters which
constitute single-age, single-metallicity and equal-distance stellar
samples. Fig.~\ref{fig_kroupa:CLlf} does exactly this, and indeed, a
very pronounced peak is evident at exactly the right location and with
the right width and height (Kroupa 2002: fig.~1).
We can therefore trust the peak in $\Psi_{\rm phot}$. In
Fig.~\ref{fig_kroupa:MWlf} it can be seen that $\Psi_{\rm near}$ also
shows evidence for this peak, noting that the peak is smeared apart in
the local stellar sample because the stars have a wide spread in
metallicities. The metallicity-dependence of the peak has been shown
to be in agreement with the LFs of star clusters over a large range of
metallicity (von Hippel et al. 1996; Kroupa \& Tout 1997).

Kroupa, Tout \& Gilmore (1993) then performed a trick to get at the
correct mass--luminosity relation without having to resort to
theoretical or purely empirical relations which are very uncertain
in
 their derivatives (Kroupa 2002: fig.~2): The Malmquist-corrected
$\Psi_{\rm phot}$ is used to
 define the amplitude and width and
location of the extremum in the single-age, single-metallicity average
$dm/dM_V$, and integrating the resulting constraint leads to a
semi-empirical $m(M_V)$ relation. It is a semi-empirical relation
because we used theoretical stellar models to place, one can say,
zeroth order constraints on this relation, i.e.  to prove the
existence of the extremum and to estimate its location and width and
amplitude.  With this theoretical knowledge in-hand we then used the
LF to constrain the detailed run of $m(M_V)$. The result is in
amasing
 agreement even with the most recent high-quality
binary-star
 constraints by Delfosse et al. (2000). With the
so-obtained $m(M_V)$
 relation, which has the correct derivative, it
is now possible to
 take another step towards constraining the MF. 

But first the problem $\Psi_{\rm near} \ne \Psi_{\rm phot}; M_V > 12$
needs to be addressed.

\subsection{Unresolved binary systems}

\noindent One bias clearly affecting $\Psi_{\rm phot}$ are unresolved
multiple systems; in constructing $\Psi_{\rm near}$ all stellar
companions have been counted individually, while $\Psi_{\rm phot}$
consists of ``stars'' too far away to be resolved if they are
multiple. Also, a faint companion star may not be seen because it lies
below the flux limit while the primary enters the formal photometric
distance limit. Furthermore, an unresolved binary system appears
redder unless both stars are of equal mass and this affects the
calculated stellar space density because the photometric distance is
misjudged. All the effects can and must be corrected for; this having
been done thoroughly for the first time by Kroupa et al. (1993).

The bias due to unresolved binaries affecting $\Psi_{\rm phot}$ can
be
 nicely demonstrated using realistic numbers for the multiplicity
in
 the Galactic field (Goodwin et al. 2006): Suppose the observer
sees
 100 systems on the sky. Of these 40 are binaries, 15 triples
and 5 quadruples. The multiplicity fraction is $f=(40+15+5)/100 =
0.6$.  The observer will construct the LF using these 100 systems, but
in fact
 85 stars are missed. All of these are much fainter than
their primary, implying
 a non-linear depression at the faint end of
the LF. 

\subsection{The standard Galactic-field IMF}

\noindent
 Together with a thorough modelling of the star-formation
history of
 the solar neighbourhood (low-mass stars take long to
descend to the
 main sequence), local Galactic disk structure and
metallicity spreads,
 as well as different fractions of multiple
stars and photometric and
 trigonometric distance uncertainties,
Kroupa et al. (1993) performed a
 multi-dimensional optimisation for
the parameters of a two-part-power
 law MF. The result is one MF
which unifies both $\Psi_{\rm near}$ and
 $\Psi_{\rm phot}$. It is
given by eq.~\ref{eq_kroupa:canonIMF} for
 $m\simless 1\,M_\odot$
(Kroupa et al. 1993; Reid et
 al. 2002). Adopting this IMF as being
correct, Kroupa (1995b) also
 showed that the observed LFs are
arrived at if stars are born as binaries in clusters which evolve and
disperse their stellar
 content into the Galactic field. Chabrier
(2003) further points out
 that non-linearities in the
colour--magnitude relation used in
 photometric parallax adds to the
underestimate of the stellar
 densities at the faint end of
$\Psi_{\rm phot}$.

Scalo (1986) had performed a very detailed analysis of the stellar
mass function, which was superseded for $m\simless 1\,M_\odot$ through
new star-count data and the new modelling described above. But for
$m\simgreat 1\,M_\odot$ his analysis remains valid. Using star counts
out to kpc distances for early-type stars he modelled their spatial
distribution and took account of their production through different
star-formation rates, and arrived at an estimate of the Galactic-field
IMF adopted in Kroupa et al. (1993):
\begin{equation}
\begin{array}{l@{\quad\quad,\quad}r@{\;}l}
\alpha_3 = 2.7\pm0.7,    &1 \simless  &m/M_\odot.
\end{array}
\label{eq_kroupa:ScaloIMF}
\end{equation}
Together with eq.~\ref{eq_kroupa:canonIMF} this is the {\it standard
Galactic-field IMF}.  Elmegreen \& Scalo (2006) then showed that
there can be artificial features in the massive-part of the
Galactic-field IMF when it is deduced from the present-day field MF
under varying star-formation rates (SFRs), frustrating
 attempts to
attribute any possible structure there to star-formation
 physics.

\subsection{The canonical IMF}
\label{sec_kroupa:canIMF}

\noindent
As already stated in \S~\ref{sec_kroupa:intro}, the IMF for early-type
stars can also be inferred for individual OB associations and star
clusters. Massey (2003 and previous papers) has shown the IMF for
massive stars to be independent of density ($\rho_{\rm central}
\simless 10^{5.9}$~stars/pc$^3$, the densest cluster being R136 in the
LMC) and metallicity ($Z\simgreat 0.002$, the population with the
lowest metal abundance being in the SMC). It always has
$\alpha=2.3\pm0.1$, thus leading to the standard or canonical stellar
IMF, eq.~\ref{eq_kroupa:canonIMF}.

The question now emerges as to why $\alpha_{\rm Scalo} = 2.7 >
\alpha_{\rm Massey}=2.3$ for $m\simgreat 1\,M\odot$. This is referred
to as the ``Scaslo {\sl vs} Massey discrepancy''.  

Considering the uncertainties, it would be valid to discard the
difference. However, using the standard Galactic-field IMF in
galactic
 evolution models would yield about~3 times fewer stars in
the mass range $10-150\,M_\odot$ and about 8 times fewer stars with
$150\,M_\odot$ than when using the canonical IMF. So it is important
to know $\alpha_3$. We will return to this later-on, but as before,
when not immediately
 seeing a possible solution it often proves
useful to ignore the
 problem until new insights open new avenues for
exploration.

\subsection{Massive stars come seldomly alone}
\label{sec_kroupa:massive_bins}

\noindent
 Spectroscopic and speckle-interferometric observations of
nearby
 massive star-forming regions have shown that OB stars prefer
more than
 one partner (Zinnecker 2003; Goodwin et al. 2006). In
particular, Preibisch et al. (1999) present the most
 detailed
multiplicity study of the Orion Nebula Cluster (ONC)
 demonstrating
that the rule are triples rather than binaries as
 amongst T~Tauri
stars.

This allows us to set-up the following {\it hypothesis}: 
\begin{equation}
{\alpha_{\rm Massey} \over \alpha_{\rm Scalo}} < 1 \quad {\rm because}\quad
f_{\rm Scalo}<1 \quad {\rm while} \quad f_{\rm Massey} \simgreat 1.
\label{eq_kroupa:hyp}
\end{equation}
That is, Massey was observing dynamically less-evolved and therefore
more binary-rich populations whereas Scalo saw the field population
which is, to a significant extend, ``contaminated'' by runaway OB
stars which are mostly single (Stone 1991). This would be a neat
solution to our problem: fainter OB stars are hidden among their
brighter primaries flattening the Massey-IMF, while the mostly
single
 OB-field stars in the Scalo sample would be a more correct
representation of the true IMF.
 
 Thus we may be led to conclude
that in actuality the true
 (binary-corrected) IMF is
eq.~\ref{eq_kroupa:ScaloIMF}
 ($\alpha_3\approx 2.7$), while
eq.~\ref{eq_kroupa:canonIMF}
 ($\alpha_3\approx 2.3$) is wrong.  This
would, of course, have
 important implications for Galactic
astrophysics as well as cosmology; most galaxy-formation and evolution
calculations are being done with
 a Salpeter IMF.

\section{Lessons from star clusters}

\noindent
Star clusters pose star-formation events which are extremely well
correlated in space, time and chemistry, and so studying the MF within
clusters avoids most of the problems we need to deal with in the
Galactic field. However, there are a number of serious issues that
essentially annihilate any advantage one might have gained by having an
equal-age, -distance and -metallicity population:

\subsection{Young clusters} 

\noindent
To avoid dynamical evolution complicating the
star-counts in a cluster, young clusters are chosen. However,
uncertainties in pre-main sequence tracks make the calculation of
stellar masses from stellar luminosities and colours or spectra very
uncertain for ages less than a few~Myr.  Errors of about 50~\% would
not be untypical. These errors are not randomly distributed though,
but cause systematic offsets from the true stars. And so unphysical
structures appear in the MF. This is discussed in more depth in Kroupa
(2002). Very young clusters also have a high binary fraction, which
again would require corresponding corrections. Young clusters evolve
very rapidly during the first~1~Myr as a result of residual gas
expulsion. For example, the ONC with an age of about
1~Myr is already between~5 and 15~initial crossing times old, implying
a rapid dynamical evolution of the binary population (fig.~1 in Kroupa
2000; fig.~9 in Kroupa et al. 2001) {\it and} an expansion of the
cluster such that a substantial fraction of its initial population has
probably been lost (Kroupa, Aarseth \& Hurley 2001). This translates
into a selective stellar-mass loss {\it if} the cluster was initially
significantly mass-segregated (Moraux, Kroupa \& Bouvier 2004).  It is
therefore never clear exactly which corrections have to be applied, and
an inferred stellar MF is very likely not to be a good representation
of the IMF.

\subsection{Old clusters}

\noindent
To avoid the issues above (uncertain pre-main sequence theory, rapidly
evolving stellar and binary populations), older clusters within which
stars are on or close to the main sequence would appear to be
useful. They are definitely useful, but such clusters are heavily
evolved dynamically, and so the MF will not represent the IMF even for
clusters with an age near~100~Myr. Unresolved multiple systems remain
a problem, and for example for the Pleiades Stauffer (1984) shows that
the fraction of photometric binaries is 26~\% (these are stars that
are more luminous than a single star of the same colour), while
Kaehler (1999) notes that a true binary fraction between~60 and 70~\%
may be possible.  Even globular clusters appear to have a sizeable
binary population (e.g. Rubenstein \& Bailyn 1997).

A star cluster which has managed to survive its initial gas
expulsion
 phase re-virialises on a time-scale of tens of Myr (Kroupa
et
 al. 2001). It's binary population is depleted by then such that
binaries with orbital Kepler velocities smaller than the velocity
dispersion
 in the pre-expansion cluster will have been
disrupted. The remaining
 binaries are mostly inert; sometimes
energetic binary--single-star or
 binary--binary encounters near the
cluster core eject 1--3 stars
 (e.g. a single star and the binary)
from the cluster.  The cluster's
 evolution is, however, well
described by classical dynamical evolution
 tracks, i.e. the
surviving binaries are dynamically not important, as
 shown
explicitly by Kroupa (1995c). During this long-lived phase the
cluster evolves towards energy equipartition, which implies that the
low-mass stars gain energy and thus move outwards becoming lost to
the
 Galactic tide, while the more massive stars loose energy and
segregate
 towards the center of the potential. This process never
stops, i.e. a
 cluster never actually reaches energy equipartition
and dynamical
 equilibrium. The consequence is that as the cluster
ages, it's MF
 becomes increasingly depleted in low-mass stars. This
is nicely
 evident in low-mass clusters in terms of a flattening of
the LF, while
 unresolved binaries remain a significant bias despite
their dynamical
 unimportance (fig.~10 in Kroupa 1995c).  Baumgardt
\& Makino
 (2003) study the evolution of the MF in much detail for
massive star
 clusters without binaries. In their fig.~7 they nicely
show how the MF
 changes its slope in dependence of the cluster's
dynamical age
 expressed in terms of the fraction of the time until
cluster
 disruption: clusters that have passed 50~\% of their
disruption time
 have essentially a flat ($\alpha\approx0$) MF for
$0.1\simless
 m/M_\odot \simless 0.8$, while for older clusters
$\alpha$ becomes
 negative. 

\subsection{Massive stars in clusters}

\noindent
Two processes compete for all stellar masses, but are particularly
pronounced for massive stars because of the involved energetics. 

{\bf Mass segregation:} If massive stars form throughout a forming
star cluster energy equipartition will force them to segregate to
the
 center within a timescale $t_{\rm msgr} \approx 2\,(m_{\rm av}
/
 m_{\rm massive})\,t_{\rm relax}$, where $m_{\rm av}, m_{\rm
massive}$
 are the masses of the average star and the massive stars,
and $t_{\rm
 relax}$ is the median two-body relaxation time, as
detailed in Kroupa
 (2004).  For example, for the ONC, $t_{\rm relax}
\approx 0.6$~Myr and
 $m_{\rm av}/m_{\rm massive} \approx 10^{-2}$
such that $t_{\rm
 msgr}\approx 0.1$~Myr $\ll$~age of the ONC. No
wonder that the ONC
 sports a beautiful Trapezium, although it may
also have formed at the
 centre (Bonnell \& Davies 1998).

{\bf Core decay:} The massive stars at the cluster centre, which may
have been born there or segregated there, form a short-lived
small-$N_m$ core.  It decays on a time-scale $t_{\rm decay} \approx
N_m \times t_{\rm cross,core}$, where $N_m$ is the number of massive
stars in the core and $t_{\rm cross, core}$ is its crossing time.
Again, for the ONC, we find (Kroupa 2004) $t_{\rm decay}\approx
10^4-10^5$~yr $\ll$ its age. So why does the Trapezium still exist?

\subsection{A highly abnormal ``IMF'' in the Orion Nebula Cluster}

\noindent 
 Detailed $N$-body computations confirm this time-scale
problem
 (Pflamm-Altenburg \& Kroupa 2006): The ONC also has a
highly
 abnormal ``IMF'' - only 10 stars more massive than
$5\,M_\odot$ are
 found in it, while 40 would be needed to allow the
very final remnant
 of the central core to be still visible today
{\it and} to account for the
 mass of the most massive star, given
the cluster-mass--maximal-mass
 relation observed to exist among
young clusters (Weidner \& Kroupa 2006).

This appears to be the rule rather than the exception,
Pflamm-Altenburg \& Kroupa (2006) discuss the older Upper Scorpius
OB
 association which also shows a deficit of massive stars compared
to
 the canonical IMF. {\it Naively this may be taken as good
evidence for a bottom-heavy present-day IMF in a relatively metal-rich
environment}, which would even be consistent
 qualitatively with the
theoretical arguments of
 \S~\ref{sec_kroupa:theory}
(higher-metallicity environments making
 lower-mass stars on
average).

Further work, however, shows this to be an illusion: Pflamm-Altenburg
\& Kroupa (2006) apply a specially designed chain-regularisation code
({\sc CATENA}) to study the dynamical stability of ONC-type cores,
finding that if the ONC and Upper Scorpius OB association, which is
understood to be an evolved version of the ONC, were born with 40 OB
stars, then the observations are well-accounted for. Indeed, Stone
(1991), among others, shows that 46~\% of all O~stars are probably
runaways of which about 18~\% have line-of-sight velocities $\simgreat
30$~pc/Myr. Energetic dynamical encounters in cluster cores
therefore have a very significant effect on the shape of the ``IMF''
in clusters and OB associations. {\it And}, massive stars are rapidly
dispersed from their birth sites with corresponding implications for
the spreading of heavy elements and feedback-energy deposition
throughout a galaxy.

The implications of these studies is therefore that the massive-star
content of star clusters and OB associations is not a measure of the
true initial content, unless the full dynamical history of the
cluster
 or OB association is taken into account.

\subsection{Summary}
\label{sec_kroupa:pessimism}

\noindent
 It would thus appear that clusters and OB associations of
any age are a {\it horrible}
 place to study the IMF.

This does not mean, however, that these systems are useless. Quite on
the
 contrary: they are indeed our only samples of stars formed
together
 and are therefore the only samplings from the IMF that
exist before
 dispersion into the field.  The point of the above
pessimistic note is
 to instill the fact that any measurement of the
``IMF'' in a cluster or association {\it must} be accompanied by a
dynamical investigation {\it before}
 conclusions about the possible
variation of the IMF can be made. An
 excellent example follows in
\S~\ref{sec_kroupa:GalC}.

\section{Evidence for variation of the IMF}

\noindent The derivation of MFs in many clusters and associations
leads to many
 values of $\alpha$ measured over different stellar
mass ranges.  The
 alpha-plot, $\alpha(m)$, can therefore give us
information about the
 true shape of the putative underlying IMF, and
also of the scatter
 about a mean IMF (Scalo 1998). This scatter can
then be studied
 theoretically using the methods mentioned above,
i.e. $N$-body
 calculations of star-clusters with different
properties. Thus, \S~\ref{sec_kroupa:pessimism} needs to be invoked in
the context of some recent interesting observational studies that
suggest the IMF to vary at low masses (e.g. Oasa et al. 2006) and at
high masses (Gouliermis et al. 2005).  

One important aspect here is the purely statistical scatter
resulting
 from finite-$N$ sampling from a putative universal IMF
(\S~\ref{sec_kroupa:intro}).  Elmegreen
 (1997, 1999) showed that the
observed scatter is consistent with the
 variations of $\alpha(m)$ as
a result of sampling from the
 IMF. Stellar-dynamical and unresolved
binary-star biases lead to systematic deviations of $\alpha(m)$ away
from the true value, and Kroupa (2001, 2002) has
 shown the observed
alpha-values to be consistent with the canonical
 IMF
(eq.~\ref{eq_kroupa:canonIMF}) for a large variety of stellar
populations using theoretical models of initially binary-rich star
clusters. 

As already deduced with some force by Massey (2003, and
 previous
papers), there is no evidence for systematic shifts of
 $\alpha(m)$
with metallicity nor density.  At the one extreme end of the scale of
physical conditions, globular clusters have very similar MFs as
young Galactic clusters and the Galactic field, while constraints on
the massive-star content of globular clusters are such that the IMF
could not have been too top heavy as otherwise the mass-loss from
evolving stars would have unbound the clusters (Kroupa 2001). At the
other extreme end, the super-solar Arches cluster near the Galactic
centre has a massive-star IMF again very similar to the
Salpeter/Massey one (\S~\ref{sec_kroupa:GalC} below).

Kroupa (2002) studied the distribution, $f_\alpha(\alpha)$, of
measured $\alpha$ values for stars more massive than a few~$M_\odot$
finding it to be describable by two Gaussians, one somewhat broad
but
 with symmetric wings, and the other exceedingly narrow, and both
centered on the Salpeter/Massey value $\alpha=2.35$. This is very
surprising because the various $\alpha(m)$ values were obtained
 by
different research groups observing different objects. An equivalent
theoretical data set shows a somewhat broader unsymmetric
distribution
 offset from the input Salpeter/Massey value.  This
needs further
 investigation, because one would expect the
theoretical data to be
 better ``behaved'' than the observational
ones which include the
 entire complications of inferring masses from
observed luminosities.  Concerning this open question, Ma{\'{\i}}z
Apell{\'a}niz \& {\'U}beda (2005) point out that binning the
star-counts into mass-bins of equal mass leads to biases that distort
the shape of the IMF when the number of stars is small, which is often
the case at high masses.

Returning to the empirically-determined existence of an upper mass
limit to stars near $150\,M_\odot$ (\S~\ref{sec_kroupa:intro}), it
remains to be understood why it does not seem to depend on
metallicity: Theoretical concepts would suggest that feedback
 ought
to play a role in limiting the masses of stars, but this
 implicitly
contains a metallicity dependence as the coupling of
 radiation to
the accreting gas is more efficient in metal-rich environments. Thus,
we are left yet again with two open questions.

\subsection{Back to the ``Scalo {\sl vs} Massey discrepancy''}
\label{sec_kroupa:sc_mas}

\noindent 
 One important implication from this analysis of
$\alpha(m)$ values for
 massive stars is that the above hypothesis
(eq.~\ref{eq_kroupa:hyp})
 must be discarded: if unresolved multiple
systems were the origin of
 the {\it Scalo vs Massey} discrepancy,
then $f_\alpha(\alpha)$ would
 have an asymmetric distribution
stretching from the single-star value
 ($\alpha=2.7$) to the
unresolved binary-star value
 ($\alpha=2.3$). This is not the
case. The Scalo vs. Massey discrepancy therefore is either
non-existing (perhaps
 because Scalo's analysis requires an update),
or it has another hitherto
 unknown origin. The proposition by
Elmegreen \& Scalo (2006) that the IMF calculated from the present-day
field MF may be steeper than the canonical IMF if the field was
created during a lull in the star-formation rate deserves some
attention in this context.

\subsection{An example: The Galactic central region}
\label{sec_kroupa:GalC}

\noindent
Reports do exist that in extreme environments there is evidence for a
top-heavy IMF.  These often come from indirect arguments based on
luminosity and available mass for star formation in star-bursts, for
example: scaling the canonical IMF to the observed luminosity would
lead to a mass in stars far superior to dynamical mass constraints
from kinematics, requiring either a top-heavy (smaller $\alpha_3$)
IMF, or an IMF truncated typically near or above $1\,M_\odot$. Paumard
et al. (2006) report evidence for a top-heavy IMF in both
circum-nuclear disks in the MW. While the observations are quite
convincing, \S~\ref{sec_kroupa:pessimism} needs to be invoked before
taking such results as affirmative evidence for a variable stellar
IMF.

A case in point is the Arches cluster at the Galactic centre, where
the conditions
 for star formation are very different to those
prevalent in the solar
 vicinity or the disk of the LMC, by being in
general warmer ($\simgreat 30$~K
 instead of 10~K) and having a
super-solar metallicity ($-0.15
 \simless\;$[Fe/H]$\;\simless
+0.34$).

Klessen, Spaans, Jappsen (2006) perform an excellent state-of-the art
calculations
 of star-formation under such conditions. They state in
their abstract:
 ``In the solar neighbourhood, the mass distribution
of stars follows a
 seemingly universal pattern. In the centre of the
Milky Way, however,
 there are indications for strong deviations and
the same may be true
 for the nuclei of distant star-burst
galaxies. Here we present the
 first numerical hydro-dynamical
calculations of stars formed in a
 molecular region with chemical and
thermodynamic properties similar to
 those of warm and dusty
circum-nuclear star-burst regions. The resulting
 initial mass
function is top-heavy with a peak at $\approx
 15\,M_\odot$, a sharp
turn-down below $\approx 7\,M_\odot$ and a
 power-law decline at high
masses. We find a natural explanation for
 our results in terms of
the temperature dependence of the Jeans mass,
 with collapse
occurring at a temperature of $\approx 100$~K and an
 H$_2$ density
of a few $10^5$~cm$^{-3}$, ...''. Theirs would be in
 agreement with
the previously reported apparent decline of the stellar
 MF in the
Arches cluster below about $6\,M_\odot$.

Shortly after the above theoretical study appeared, Kim et
al. (2006)
 published their observations of the Arches cluster and
performed the
 necessary state-of-the art $N$-body calculations of
the dynamical
 evolution of this young cluster, revising our
knowledge
 significantly. Quoting from their abstract: ``We find that
the
 previously reported turnover at $6\,M_\odot$ is simply due to a
local
 bump in the mass function (MF), and that the MF continues to
increase
 down to our 50~\% completeness limit ($1.3\,M_\odot$) with
a power-law
 exponent of $\Gamma = -0.91$ for the mass range of $1.3
< m/M_\odot <
 50$. Our numerical calculations for the evolution of
the Arches
 cluster show that the $\Gamma$ values for our annulus
increase by 0.1-0.2
 during the lifetime of the cluster, and thus
suggest that the Arches
 cluster initially had a Gamma of $-1.0$ --
$-1.1$, which is only
 slightly shallower than the Salpeter value.''
($\Gamma = 1-\alpha$).
 
 This serves to illustrate the general
observation that when detailed
 high-resolution observations {\it
and} accurate $N$-body calculations
 are combined, claims of abnormal
IMFs tend to vanish, irrespective of
 what theory prefers to
have. Another example, the ONC, has already
 been noted above.

\subsection{Spheroids:  old, very extreme star-bursts?}
\label{sec_kroupa:bulge}

\noindent
An interesting, albeit indirect, result by Francesca Matteucci and
her
 collaborators based on multi-zone photo-chemical evolution
studies of
 the MW, its Bulge and elliptical galaxies shows the MW
disk to be
 re-producible by the standard Galactic-field IMF
(eq.~\ref{eq_kroupa:ScaloIMF}, $\alpha_3\approx 2.7$) but not the
canonical IMF (eq.~\ref{eq_kroupa:canonIMF}). This is supported by the
independent work of Portinari, Sommer-Larsen \& Tantalo (2004).  In
contrast, the MW Bulge may require a rapid phase of formation lasting
0.01--0.1~Gyr and a top-heavy IMF with $\alpha\approx1.95$ (Romano et
al. 2005;
 Ballero, Matteucci \& Origlia 2006). For elliptical
galaxies a similar
 scenario seems to hold, whereby the IMF needed to
account for the
 colours and chemical properties is the canonical one
($\alpha_3\approx
 2.3$), but flattening (decreasing $\alpha_3$)
slightly with increasing
 galaxy mass (Pipino \& Matteucci 2004).  An
experiment, where the yields for massive stars are changed as much as
possible without violating available constraints confirms this result
- it seems not to be possible to understand the metallicity
distribution of MW Bulge stars with a standard Galactic field
IMF. Instead, $\alpha_3\approx 1.9$ seems to be required (Ballero,
Kroupa \& Matteucci 2007). Nevertheless, uncertainties remain. Thus,
Samland, Hensler \& Theis (1997) find the Bulge to have formed over a
time-span of about 4~Gyr with a Salpeter IMF using a two-dimensional
chemo-dynamical code, while Immeli, Samland \& Gerhard (2004) obtain
formation time-scales near 1~Gyr using a three-dimensional
chemo-dynamical code. Finally, Zoccali et al. (2006) find the ``MW
bulge to be similar to early-type galaxies, in being $\alpha$-element
enhanced, dominated by old stellar populations, and having formed on a
timescale shorter than $\approx$1 Gyr''. ``Therefore, like early-type
galaxies the MW bulge is likely to have formed through a short series
of starbursts triggered bythe coalescence of gas-rich mergers, when
the universe was only a few Gyr old''. For low-mass stars, Zoccali et
al. (2000) find the Bulge to have a similar MF as the disk and
globular clusters.

\subsection{Composite populations}
\label{sec_kroupa:compIMF}

\noindent The ``Scalo {\sl vs} Massey discrepancy'' found in
\S~\ref{sec_kroupa:canIMF} and its proposed solution in
\S~\ref{sec_kroupa:massive_bins} in terms of unresolved multiple
systems was found, in \S~\ref{sec_kroupa:sc_mas}, not to be
possible. Again, as in \S~\ref{sec_kroupa:canIMF} we might argue
that trying to solve the discrepancy is not worth the effort, given
the uncertainties.  Indeed, the probable solution drops-out through
an entirely different line-of-thought:

It needs to be remembered that star clusters are the {\it fundamental
building blocks of galaxies} (Kroupa 2005).  Therefore, to be exact,
all the stellar IMFs in all the clusters forming in one
``star-formation epoch'' (Weidner, Kroupa \& Larsen 2004) of a galaxy
must be added to calculate the global, galaxy-wide ``integrated
galactic IMF'' (IGIMF). This has been pointed-out by Vanbeveren (1982,
1984) and discussed again by Scalo (1986), and formulated for an
entire galaxy by Kroupa \& Weidner (2003, 2005), noting that the
star-clusters are distributed as a power-law embedded-cluster mass
function with an exponent $\beta\approx 2\pm0.4$, where the Salpeter
value would be~2.35. Also, an empirical relation exists limiting the
mass of the most massive star in a cluster with stellar mass $M_{\rm
ecl}$: $m\le m_{\rm max}(M_{\rm ecl})$ (Weidner \& Kroupa 2006),
enhancing the {\it IGIMF$\ne$IMF effect}.

Putting this together, an integral over all clusters formed together
in one epoch results, and the composite IMF turns out to be steeper
with $\alpha_{\rm 3,IGIMF}>\alpha_3$ ($=\;$the canonical IMF value)
for $m\simgreat 1.6\,M_\odot$ with a steep downturn at a mass $m_{\rm
max,\,gal}$. Both, $\alpha_{\rm3, IGIMF}$ and the galaxy-wide maximum
star mass, $m_{\rm max,\,gal}$, depend on the star-formation rate of
the galaxy, and so the situation becomes complex with important
implications for galactic spectro-photometric and chemical evolution
studies given that the number of supernovae of type~II is depressed
significantly per unit stellar mass formed when compared to a
universal Salpeter IMF, which is often assumed in cosmological
applications (Goodwin \& Pagel 2004; Weidner \& Kroupa 2005). Koeppen,
Weidner \& Kroupa (2006) show that this IGIMF-ansatz naturally
explains the observed form of the mass--metallicity relation of
galaxies without the need to invoke selective outflows (which are not
excluded though).

Applied to the MW, Scalo's $\alpha_3\approx 2.7$ comes-out naturally
for an input canonical IMF. This is also required by
spectro-photometric and chemical-evolution models of the MW disk
(\S~\ref{sec_kroupa:bulge}). At the same time, for LSB galaxies the
IGIMF must be steep, i.e. bottom-heavy. Again, this seems to be
confirmed by observations (Lee et al. 2004).  Is this, then, the
solution to the ``Scalo {\sl vs}
 Massey discrepancy''?

Elmegreen (2006), however, opposes the notion that the IGIMF is
different to the stellar IMFs found in individual clusters, mostly
because the star-cluster mass function is not steep steep enough
($\beta < 2$), in which case ${\rm IGIMF}\approx {\rm IMF}$ as
already
 shown by Kroupa \& Weidner (2003). {\it Much therefore rests
on the
 detailed shape of the star-cluster initial mass function}\,!
Also, the
 existence of a physical $m_{\rm max}(M_{\rm ecl})$
relation is of
 central importance to the $IGIMF \ne IMF$ argument,
but opposing views
 are being voiced (e.g. Elmegreen 2006). An
argument against IGIMF~$\ne~$IMF is that some studies suggest these to
be equal, e.g. in elliptical galaxies
(\S~\ref{sec_kroupa:compIMF}). However, an appreciable fraction of the
research which Elmegreen fields in-support of his equality-conjecture
actually support the inequality-conjecture. So this issue is still
very much under debate and may need observational improvement by (i)
better constraining the true initial star-cluster mass function and
(ii) improving the empirical $m_{\rm
 max}(M_{\rm ecl})$
relation. Observational work is underway to test the IGIMF$\ne$IMF
notion, and for example Selman \& Melnick (2005) find the composite
IMF to be flatter than the stellar IMF in the 30~Dor star-forming
region, as predicted, but only with low-significance.

\section{Concluding remarks}

\noindent
It would have seemed that when two mutually more or less exclusive
theories of the origin of stellar masses make the same basic
prediction concerning the variation of the average stellar mass with
physical conditions, that this expected variation would be very
robust
 and born out in observational data.  Alas, the observational
data on
 the IMF are resilient - they do not yield what we desire to
see. The
 stellar IMF is invariant and can be best described by a
two-part power-law
 form (eq.~\ref{eq_kroupa:canonIMF}).  This holds
true for
 metallicities ranging from those of globular clusters to
super-solar
 values near the Galactic centre, and for densities less
than about
 $10^{6}$~stars/pc$^3$. Even the maximal stellar mass of
about
 $150\,M_\odot$ seems to be independent of metallicity for
$Z\simgreat
 0.008$.

Given the quite total failure to account for the resilience of the
stellar IMF towards changes, it is clear that this IMF-conservatism
poses some rather severe challenges on liberal star formation theory.

Only indirect arguments based on chemical evolution work seem to
suggest the IMF to have been somewhat flatter for massive stars
during
 the extreme physical conditions valid during the formation of
the MW
 Bulge and elliptical galaxies. Variations of the galaxy-wide
IMF (the
 IGIMF) among galaxies and also in time can be understood
today despite
 the existence of a universal canonical IMF, but this
issue is still
 very much in debate (Elmegreen 2006 {\sl vs} Weidner
\& Kroupa 2006).  The evidence noted in \S~\ref{sec_kroupa:bulge}
appears to suggest that perhaps $\alpha_{\rm 3,IGIMF}\approx 2.7$ for
the MW disk, $\alpha_{\rm 3,IGIMF}\approx 2.3$ for E-galaxies, while
$\alpha_{\rm 3,IGIMF}\approx 1.9$ for massive E-galaxies and the MW
Bulge. Is this true?


\newpage

\noindent
{\small {\sc Acknowledgments}: I would like to thank the organisers
sincerely for a most enjoyable meeting and my collaborators for
important contributions. This research was supported by an Isaac
Newton and a Senior Rouse Ball Studentship in Cambridge in the UK, and
by the Heisenberg programme of the DFG and DFG research grants in
Germany. This article was mostly written in Vienna and I thank
Christian Theis and Gerhardt Hensler for their hospitality.}

\end{document}